\documentclass[aps,prb,twocolumn,amsmath,amssymb,superscriptaddress,floatfix]{revtex4}
\usepackage{graphicx}
\usepackage{bm}
\usepackage{amssymb}
\usepackage{amsmath}
\usepackage{color}
\newcommand{\be}{\begin{equation}}
\newcommand{\ee}{\end{equation}}
\newcommand{\beqn}{\begin{eqnarray}}
\newcommand{\eeqn}{\end{eqnarray}}

\begin{document}

\title{Transverse-spin correlations of the random transverse-field Ising model}

\author{Ferenc Igl\'oi}
\email{igloi.ferenc@wigner.mta.hu}
\affiliation{Wigner Research Centre for Physics, Institute for Solid State Physics and Optics,
H-1525 Budapest, P.O.Box 49, Hungary}
\affiliation{Institute of Theoretical Physics,
Szeged University, H-6720 Szeged, Hungary}
\author{Istv\'an A. Kov\'acs}
\email{kovacs.istvan@wigner.mta.hu}
\affiliation{Wigner Research Centre for Physics, Institute for Solid State
Physics and Optics, H-1525 Budapest, P.O.Box 49, Hungary}
\affiliation{Institute of Theoretical Physics, Szeged University, H-6720 Szeged,
Hungary}
\affiliation{Center for Complex Networks Research and Department of Physics,
Northeastern University, 177 Huntington Avenue,
Boston, MA 02115, USA}

\date{\today}

\begin{abstract}
The critical behavior of the random transverse-field Ising model in finite dimensional lattices is governed by infinite disorder fixed points, several properties of which have already been calculated by the use of the strong disorder renormalization group (SDRG) method. Here we extend these studies and calculate the connected transverse-spin correlation function
by a numerical implementation of the SDRG method in $d=1,2$ and $3$ dimensions. 
At the critical point an algebraic decay of the form $\sim r^{-\eta_t}$ is found, with a decay exponent being approximately $\eta_t \approx 2+2d$. In $d=1$ the results are related to dimer-dimer correlations in the random AF XX-chain and have been tested by numerical calculations using free-fermionic techniques.
\end{abstract}

\maketitle
\section{Introduction}
\label{sec:intr}
The critical properties of random quantum systems with short-range interactions at $T=0$ temperature are often controlled by infinite disorder fixed points\cite{danielreview} (IDFP-s). Well known examples are the random transverse-field Ising model\cite{fisher,2d,2dRG,ddRG} (RTIM), the random antiferromagnetic (AF) Heisenberg chain\cite{fisher_XX}, but also non-equilibrium models, such as the random contact process\cite{hiv,vd} has an IDFP. The scaling properties of the IDFP-s are most conveniently studied by a special real-space renormalization group method, which is often called as strong disorder renormalization group\cite{im} (SDRG). This method has been generalized to study the scaling properties of excited states (RSRG-X method), as well as non-equilibrium dynamical properties of such systems\cite{pekker,Vosk_13,Vosk_14,cecile}. In the traditional SDRG method one constructs the ground state (and the first excited states) iteratively, by eliminating successively the local degrees of freedom having the largest excitation energy, thus the smallest time-scale. Then, between the remaining degrees of freedom new terms are generated in the renormalized Hamiltonian. At an IDFP the terms in the Hamiltonian (couplings, transverse-fields, etc.) have a very broad distribution, so that the strength of disorder tends to infinity. At this limit the perturbative treatment of the calculation of the renormalized terms becomes asymptotically exact, so that the SDRG is expected to provide correct critical properties in the scaling limit.

The prototype of random quantum magnets is the RTIM, which is defined by the Hamiltonian:
\be
{\cal H} =
-\frac{1}{2}\sum_{\langle ij \rangle} J_{ij}\sigma_i^x \sigma_{j}^x-\frac{1}{2}\sum_{i} h_i \sigma_i^z\;,
\label{eq:H}
\ee
where the $\sigma_i^{x,z}$ are Pauli-matrices and $i$, $j$ denote sites of the lattice. In Eq.(\ref{eq:H}) ${\langle ij \rangle}$ refers to
nearest neighbours and the $J_{ij}$ couplings and the $h_i$ transverse fields are independent random numbers,
which are taken from the distributions, $p(J)$ and $q(h)$, respectively. In the following we restrict ourselves to ferromagnetic, non-frustrated models and define the control-parameter as $\theta=\overline{\ln h}-\overline{\ln J}$. Here and in the following we denote by $\overline{x}$ the average value of $x$ over quenched disorder. The critical point of the system is denoted by $\theta_c$, so that for $\theta>\theta_c$ ($\theta<\theta_c$) the system is in the paramagnetic (ferromagnetic) phase.

The critical properties of the RTIM are thoroughly studied by the SDRG approach and by other exact and numerical methods\cite{young_rieger,igloi_rieger,pich,matoz_fernandez}. In $d=1$ we have a set of presumably exact results, many of those have been obtained by Fisher through analytical solution of the SDRG equations\cite{fisher}. In $d=1$ the critical point is at $\theta_c=0$ and the average longitudinal spin-spin correlation function:
$G(r)=\overline{\left<\sigma_i^x \sigma_{i+r}^x\right>}$ scales as $G(r) \sim \exp(-r/\xi) r^{-\eta}$. Here $\left< \cdots \right>$ denotes the average value in the ground state. The average correlation length is divergent as $\xi \sim |\theta|^{-\nu}$, with $\nu=2$. At the critical point the decay of $G(r)$ is algebraic
with an exponent $\eta=(3-\sqrt{5})/2$. Between the time-scale, $\tau$, and the length-scale, $\xi$ there is a logarithmic relation: $\ln \tau \sim \xi^{\psi}$, with an exponent: $\psi=1/2$. These results have been confirmed by numerical calculations using free-fermionic techniques\cite{young_rieger,igloi_rieger}.

In higher dimensions the critical properties of the RTIM have been studied by numerical implementation of the SDRG method\cite{2d,2dRG,ddRG}. Using a very effective numerical algorithm\cite{2dRG,ddRG} it has been shown that in finite-dimensional lattices the critical behavior of the RTIM is governed by IDFP-s, at which the scaling behavior is analogous to that found in $d=1$. The set of critical exponents: $\nu,\eta,\psi$ has been calculated numerically\cite{2dRG,ddRG} for $d=2,3$ and $4$, and these were found to be independent of the actual form of disorder, thus do not depend on the distributions $p(J)$ and $q(h)$.

Besides the critical parameters which have been calculated for the RTIM so far there are still other physical observables which are needed to be investigated. One such quantity is the average transverse-spin correlation function defined by:
\be
{G_t}(r)=\overline{\left<\sigma^z_i \sigma^z_{i+r}\right>}-\overline{\left<\sigma^z_i\right>}.\overline{\left<\sigma^z_{i+r}\right>}\;.
\label{G_t}
\ee
Note, that $\overline{\left<\sigma^z_i\right>}>0$, which is proportional to the local energy-density, therefore in Eq.(\ref{G_t}) we consider the connected correlation function. At the critical point we expect an algebraic decay:
\be
{G_t}(r) \sim r^{-\eta_t}\;,
\ee
and our aim is to calculate the decay exponent $\eta_t$ in different spatial dimensions. In $d=1$ the SDRG equations of the RTIM are analogous to that of the random AF $XX$-chain\cite{fisher_XX} and there is an exact mapping between the physical observables in the two systems\cite{pst,IJR00,IJ07}. Therefore we also aim to uncover the analogous correlation function of the random $XX$-chain.

The structure of the rest of the paper is the following. The essence of the SDRG method to calculate transverse-spin correlations is presented in Sec.\ref{sec:SDRG}. The one-dimensional model is studied in details in Sec.\ref{sec:1d}, including SDRG treatment, free-fermionic calculations and mapping to the random AF XX-chain. The model in 2d and 3d is studied numerically in Sec.\ref{sec:2d3d}. Our results are discussed in Sec.\ref{Sec:disc} and details of the mapping between the 1d RTIM and the random AF $XX$-chain is presented in the Appendix.

\section{SDRG study of transverse-spin correlations}
\label{sec:SDRG}

During the renormalization procedure at each step the largest term of the Hamiltonian in Eq.(\ref{eq:H}) (denoted by $\Omega$) is eliminated and new terms are generated through a second-order perturbation calculation between the remaining degrees of freedom. If the largest term is a coupling, say $J_{ij}=\Omega$, then the two connected sites, $i$ and $j$ are coupled to form a cluster. This spin-cluster then perceives an effective transverse field of strength $\tilde{h}_{ij}\approx \frac{h_i h_j}{J_{ij}}$. If the largest term is a transverse field, say $h_i=\Omega$, then this site has negligible contribution to the (longitudinal) susceptibility, therefore decimated out. At the same time new effective couplings are generated between all sites, say $j$ and $k$, which were nearest neighbours of $i$. The effective couplings are given by: $\tilde{J}_{jk}\approx \frac{J_{ij} J_{ik}}{h_{i}}$. In higher dimensions the topology of the lattice is modified during renormalization and often two couplings are present between two sites. In this case instead of their sum, their maximum value is used as the renormalized coupling. This so called \textit{maximum rule} is justified at an IDFP, furthermore the numerical algorithms are more efficient if the maximum rule is used. (For details see in Refs.\onlinecite{2dRG,ddRG}.)

In practical applications of the SDRG a large finite sample is renormalized up to the last effective spin and the original sites of the sample are parts of effective clusters of different sizes. In the \textit{paramagnetic phase} the clusters have a finite typical linear extent, which characterizes the correlation length of the system. On the contrary in the \textit{ferromagnetic phase} there is a huge connected cluster, which is compact and contains a finite fraction of sites, $m$, being the average longitudinal magnetization. At the \textit{critical point} the giant cluster is a fractal, its total moment, $\mu$ scales with the linear size of the system, $L$ as $\mu \sim L^{d_f}$, $d_f$ being the fractal dimension. This is related to the scaling dimension of the longitudinal magnetization as $x=d-d_f$.

In the RTIM the \textit{transverse magnetization} is also position dependent.
In the SDRG process it has a large ($\sim 1)$ value at such sites of the sample, where the transverse-field satisfies the condition $h_i=\Omega$, thus these sites are separated one-spin clusters. At any other sites which belong to larger (at least two-spin) clusters the transverse magnetization is negligible. Consequently the average transverse magnetization, $m_t$, is given by the fraction of sites being one-spin clusters. As discussed in the introduction the average transverse magnetization, is non-zero, $m_t=\overline{\left<\sigma^z_i\right>}>0$, also at the critical point. Therefore we consider the connected average transverse-spin correlation function, as defined in Eq.(\ref{G_t}).

\section{Transverse-spin correlations in 1d}
\label{sec:1d}

\subsection{Numerical SDRG results}
\label{sec:SDRG1d}

\begin{figure}
  \begin{center}
    \includegraphics[width=10cm]{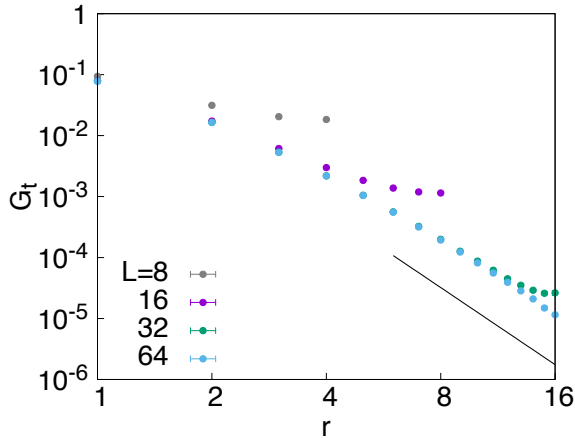}
  \end{center}
  \caption{Average connected transverse-spin correlation function of the RTIM in $1d$ calculated by the numerical implementation of the SDRG method. In the log-log plot the straight line has the slope of $-4.2$, which agrees to the estimated asymptotical value of the decay exponent as shown in Fig. \ref{Fig5}.}
  \label{Fig1}
\end{figure}

We start our analysis with the RTIM in 1d and consider periodic chains of length $L \le 64$ and the distance between the two sites being $r \le L/2$. For the actual form of disorder we have used uniform distributions: $p(J)=\Theta(J)\Theta(1-J)$ ($\Theta(x)$ being the Heaviside step-function) and $q(h)=\dfrac{1}{h_b}\Theta(h)\Theta(h_b-h)$. (In Refs.\onlinecite{2dRG,ddRG} this type of distribution is called "box-distribution", and this specific form of disorder was necessary to obtain good convergence of the numerical results.) In this case the quantum control parameter is defined as $\theta=\ln(h_b)$ and its critical value is $\theta_c(1d)=0$. We have calculated the connected transverse-spin correlation function by numerical implementation of the SDRG method and averaging is performed over $\sim10^9$ disordered samples and in each sample over the possible position of the starting point, $i$. The average connected transverse-spin correlation function calculated in this way is presented in Fig.\ref{Fig1} showing a power-law decay. Due to the relatively large value of the decay exponent we should restrict ourselves up to a distance $r \le 16$ and should consider a large number of samples to reduce the statistical error. In order to obtain a precise estimate we have calculated finite-size effective exponents by two-point fits, which are than extrapolated assuming a ${\cal O}(r^{-1})$ correction term. These are presented in Fig.\ref{Fig5}, having an estimate of the exponent $\eta_t(1d)=4.1(1)$.
In the following we show, that within the SDRG approach the decay exponent, $\eta_t(1d)\approx4$ can be estimated reasonably well by a simple argument. For this purpose we estimate the fluctuation in the number of isolated one-spin clusters at a length-scale $r$ at the critical point. The largest fluctuation is due to the presence (or absence) of effective coupling between two sites of a distance $\sim r$. It is known from the analytical solution of the SDRG equations\cite{fisher,im} that the probability that a given site is active, i.e. not decimated until a length-scale $r$ is given by: $n(r) \sim 1/r$. The probability that two sites are active at the same time is $n(r)^2 \sim 1/r^2$. In later steps of the renormalization these two active sites become connected by an effective coupling with high probability, thus the largest fluctuation in the number of isolated one-spin clusters asymptotically scales as $n(r)^2 \sim 1/r^2$. Keeping in mind that the connected transverse-spin correlation function measures fluctuation-fluctuation correlations, thus it should decay as $[n(r)^2]^2 \sim 1/r^4$, which is close to the numerically measured value of $\eta_t(1d)=4.1(1)$.

\subsection{Free-fermion calculation}
\label{sec:free_fermion}

We have confronted the results obtained by the SDRG method with direct calculation of the transverse-spin correlation function using free-fermionic techniques~\cite{lsm,pfeuty}.
The spin operators
${\sigma}_i^{x,y,z}$ are expressed in terms of fermion creation (annihilation) operators
${c}_i^\dagger$ (${c}_i$) by using the Jordan-Wigner
transformation~\cite{JW}:  ${c}^\dagger_i={a}_i^+\exp\left[\pi \imath \sum_{j}^{i-1}{a}_j^+{a}_j^-\right]$
and ${c}_i=\exp\left[\pi \imath
\sum_{j}^{i-1}{a}_j^+{a}_j^-\right]{a}_i^-$, where ${a}_j^{\pm}=({\sigma}_j^x \pm \imath{\sigma}_j^y)/2$.
The Ising Hamiltonian in~(\ref{eq:H}) can then be written in a quadratic form in fermion operators:
\beqn
{\cal H}&=&-\sum_{i=1}^{L}h_i \left({c}^\dagger_i {c}_i-\frac{1}{2} \right)-\frac{1}{2}\sum_{i=1}^{L-1} J_i({c}^\dagger_i-{c}_i)({c}^\dagger_{i+1}+{c}_{i+1}) \cr
&+& \frac{1}{2}wJ_L({c}^\dagger_L-{c}_L)({c}^\dagger_{1}+{c}_{1})\;.
\label{hamiltonian1}
\eeqn
Here $w=\exp(i\pi N_c)$, with $N_c=\sum_{i=1}^{L}c_i^{\dagger}c_i$ and in the ground state $w=1$.
The quadratic Hamiltonian in Eq.(\ref{hamiltonian1}) is diagonalized through a standard Bogoliubov transformation. The correlation functions of the spin-operators in Eq.(\ref{G_t}) involve the correlation matrix:
\be
G_{m,n}=\left< (c_m^{\dagger}-c_m)(c_n^{\dagger}+c_n)\right>\;,
\label{corr_matr}
\ee
so that $\left< \sigma_i^z \sigma_{i+r}^z \right>=G_{i,i}G_{i+r,i+r}-G_{i,i+r}G_{i+r,i}$, $\left< \sigma_i^z \right>=G_{i,i}$ and $\left< \sigma_{i+r}^z \right>=G_{i+r,i+r}$.

\begin{figure}
  \begin{center}
     \includegraphics[width=10cm]{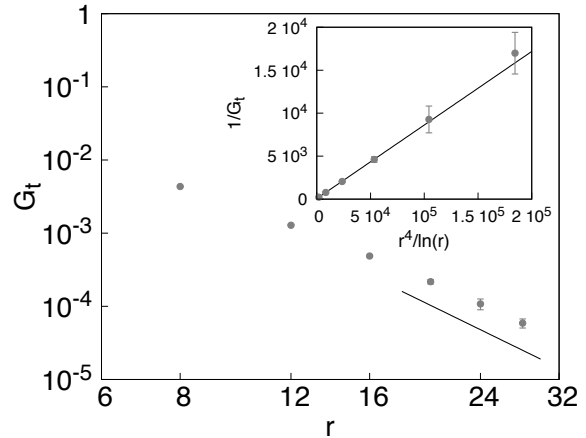}
  \end{center}
  \caption{Average connected transverse-spin correlation function of the RTIM in 1d calculated by free-fermionic methods. In the log-log plot the straight line has the slope of $-4.2$, which agrees to the estimated asymptotical value of the decay exponent as shown in Fig. \ref{Fig5}. Inset: inverse of the average connected transverse-spin correlation function of the RTIM in 1d calculated by free-fermionic methods as a function of $r^4/\ln(r)$.}
  \label{Fig2}
\end{figure}

We have calculated numerically the average connected transverse-spin correlation function of finite chains of lengths $L=16,24,32,40,48$ and $56$ having the maximal distance between the spins: $r=L/2$. The average is performed over typically $10^9$ independent random samples. In Fig.\ref{Fig2} we show $G_t(L/2)$ vs. $L$ in a log-log plot.
As seen in this figure there is an algebraic dependence: $G_t(L/2) \sim L^{-\eta_t}$, where $\eta_t$ approaches a value of $\eta_t \approx 4.3(1)$, if the finite-size corrections are assumed to have the form ${\cal O}(L^{-1})$. If, however, these corrections are logarithmic, then the appropriate scaling combination seems to be: $G_t(L/2) \sim \ln(L) L^{-4}$, which is illustrated in the inset of Fig.\ref{Fig2}.

\subsection{Relation with dimer correlations in the random AF XX-chain}
\label{sec:XX}
Here we show that the transverse-spin correlations of the RTIM in 1d are related to the dimer correlations in the random AF  XX-chain. This latter model is defined by the Hamiltonian:
\be
{\cal H_{XX}}=\sum_{i=1}^L  J_i \left( S_i^x S_{i+1}^x + S_i^y S_{i+1}^y \right)\;,
\label{hamiltonian}
\ee
in terms of the $S_i^{x,y}$ spin-$1/2$ operators. The closed chain (i.e. $J_{L+1} \equiv J_{1}$) contains $i=2, \dots, L(=\text{even})$ spins and the $J_j>0$ couplings are independent and identically distributed random variables. Two neighboring spins form a dimer and the dimer operator is defined as: $d_i^x=S_i^x S_{i+1}^x$, and similarly for $d_i^y$. Here we consider the dimer-dimer correlation function:
\be
D(r)=\frac{1}{L}\sum_{i=1}^L \left< d_i^x d_{i+r}^x \right> \;,
\label{dimer}
\ee
where we have averaged over the starting position of the first dimer. This correlation function has a finite limiting value: $\lim_{L \to \infty}\lim_{r \to \infty} D(r)=\epsilon_0^2/2$, where $\epsilon_0$ is the average ground-state energy per spin. We form the following difference:
\be
D^*(r)=\left[2D(r)-D(r+1)-D(r-1)\right](-1)^r/2 \;,
\label{dimer*}
\ee
which goes to zero for large separation.

The Hamiltonian of the XX-chain in Eq.(\ref{hamiltonian}) can be expressed as the sum of the Hamiltonians of two decoupled transverse-field Ising chains\cite{pst,IJR00,IJ07}:
\be
{\cal H_{XX}}=H_{\sigma}+H_{\tau}\;,
\label{XX_Ising}
\ee
where $\sigma^{x,z}_i$ and $\tau^{x,z}_i$ are two sets of Pauli matrices at site $i$, see details in the Appendix.

The dimer operators are expressed in terms of the Ising variables in different forms at odd $d_{2i-1}=\tau_i^z/4$ and even $d_{2i}=\sigma_i^x \sigma_{i+1}^x$ sites. Since the average dimer correlation function is independent of the parity of the reference site we consider here an odd starting position, when the average correlation function in Eq.(\ref{dimer}) is written as:
\be
D(2r)=\frac{2}{L}\sum_{i=1}^{L/2} \left< d_{2i-1}^x d_{2i-1+2r}^x \right>=
\frac{1}{2L}\sum_{i=1}^{L/2} \left< \tau_i^z \tau_{i+r}^z \right> \;.
\label{dimerI}
\ee
We have similarly:
\beqn
D(2r-1)&=&\frac{2}{L}\sum_{i=1}^{L/2} \left< d_{2i-1}^x d_{2i-2+2r}^x \right> \cr
&=&
\frac{1}{2L}\sum_{i=1}^{L/2} \left< \tau_i^z \right> \left< \sigma_{i+r-1}^x \sigma_{i+r}^x \right> \;,
\label{dimerI1}
\eeqn
where in the second equation we have made use of the fact that the two Ising chains are independent. The last term in Eq.(\ref{dimerI1}) can be written in a more transparent form in terms of dual Ising variables
$\tilde{\sigma}^z_{i-1/2}$, which are defined on the bonds of the
original Ising chain as
\beqn
\tilde{\sigma}^z_{i-1/2}=\sigma_{i-1}^x \sigma_{i}^x,\quad
\sigma_i^z=\tilde{\sigma}^x_{i-1/2} \tilde{\sigma}^x_{i+1/2}\;.
\eeqn
Thus we obtain an equivalent expression for the dimer correlation function in Eq.(\ref{dimer*}):
\beqn
D^*(2r)&=&\left[2D(2r)-D(2r+1)-D(2r-1)\right]/2 \cr
&=&\frac{1}{2L}\sum_{i=1}^{L/2} \left[ \left< \tau_i^z \tau_{i+r}^z \right> 
-\left< \tau_i^z \right> \left< \breve{\sigma}^z_{i+r}\right>\right]\;,
\label{dimerI*}
\eeqn
with $\left< \breve{\sigma}^z_{i+r}\right> \equiv \frac{1}{2}\left(\left< \tilde{\sigma}^z_{i+r-1/2}\right> +\left< \tilde{\sigma}^z_{i+r+1/2}\right> \right)$. For large-$r$ we have asymptotically $\left< \breve{\sigma}^z_{i+r}\right> = \left< \sigma^z_{i+r}\right> = \left< \tau^z_{i+r}\right>$, thus the dimer correlation function of the random AF XX-chain asymptotically equivalent to the connected transverse-spin correlation function of the RTIM. Thus we have a power-law decay:
\be
D^*(r) \sim r^{-\eta_t}\;,
\label{D_eta}
\ee
with $\eta_t\approx4$, supported by our numerical results indicating $\eta_t(1d)=4.1(1)$. Here we can repeat the SDRG scaling argument presented at the end of Sec.\ref{sec:SDRG1d} for the RTIM, which leads to an asymptotic relation: $D^*(r) \sim [G_s(r)]^2$. Here $G_s(r)$ is the average spin-spin correlation function of the random AF XX-chain, which according to SDRG results decays as $G_s(r) \sim r^{-2}$. Numerical free-fermion calculations, however, indicate the presence of a strong (possibly logarithmic) correction term\cite{IJR00}. This could imply that $D^*(r)$, as well as ${G_t}(r)$ for the RTIM might have a logarithmic correction, as analysed in the inset of Fig.\ref{Fig2}.

\section{Transverse-spin correlations in 2d and 3d}
\label{sec:2d3d}

\begin{figure}
  \begin{center}
    \includegraphics[width=10cm]{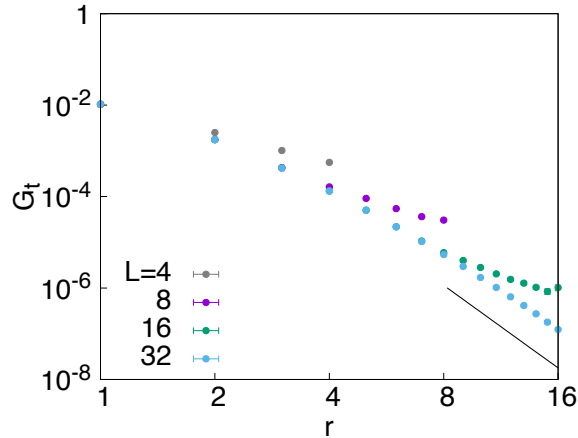}
  \end{center}
  \caption{Average connected transverse-spin correlation function of the random transverse-field Ising model in $2d$ calculated by the numerical implementation of the SDRG method. In the log-log plot the straight line has the slope of $-6$.}
  \label{Fig3}
\end{figure}

\begin{figure}
  \begin{center}
    \includegraphics[width=10cm]{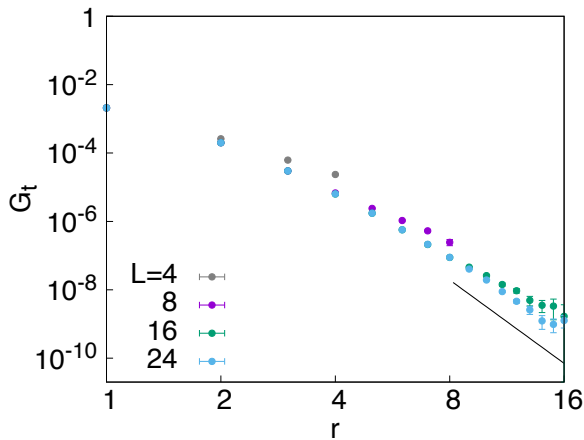}
  \end{center}
  \caption{The same as in Fig. \ref{Fig3} for the $3d$ model. In the log-log plot the straight line has the slope of $-8.1$.}
  \label{Fig4}
\end{figure}

\begin{figure}
  \begin{center}
    \includegraphics[width=10cm]{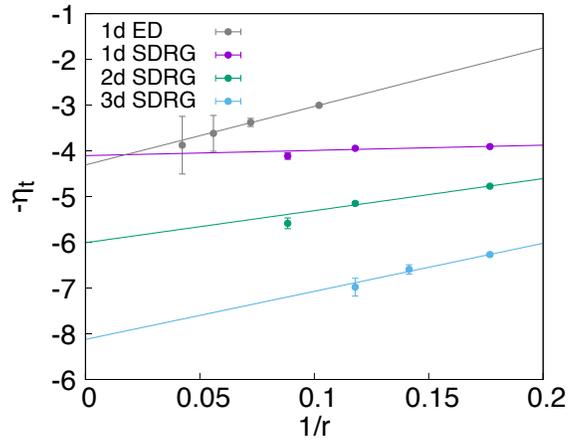}
  \end{center}
  \caption{Extrapolation of the observed finite-size $\eta_t$ values for large $r$. Finite-size estimates of $\eta_t$ are obtained through two-point fits.}
  \label{Fig5}
\end{figure}

In two- and three-dimensions we have considered the square and the simple cubic lattices, respectively, and used a very efficient numerical implementation of the SDRG algorithm as described previously in Refs.[\onlinecite{2dRG,ddRG}]. As in 1d the initial distributions of the couplings and the transverse fields were uniform, having a single parameter, $h_b$. The critical values of $h_b$ were taken from Refs.[\onlinecite{2dRG,ddRG}]. In two dimensions we have considered typically $10^9$ random samples with periodic boundary conditions having the largest linear size $L=32$. In three dimensions the number of random samples was typically $10^8$ and we went up to $L=24$. During the calculations each random sample is renormalised up to the last site and the obtained connected cluster structure is analysed using the Manhattan-distance between the sites. Correlations are averaged over all possible positions of the reference points. The calculated average connected correlations are presented in log-log scale in Figs.\ref{Fig3} and \ref{Fig4}, for $2d$ and $3d$, respectively. For both cases the average correlations are observed to follow a power-law decay.  As in 1d we have calculated finite-size effective exponents by two-point fits, which are then extrapolated assuming a ${\cal O}(r^{-1})$ correction term. These are presented in Fig.\ref{Fig5}, having estimates of the exponent $\eta_t(2d)=6.0(2)$ and $\eta_t(3d)=8.1(3)$.

\section{Discussion}
\label{Sec:disc}
In this paper we have considered a new feature of the critical behavior of the random transverse-field Ising model and calculated the connected transverse-spin correlation function. Using a numerical implementation of the SDRG approach we have observed an algebraic decay and determined the corresponding decay exponent, $\eta_t$ in one-, two- and three-dimensional lattices. The numerical results can be well approximated as $\eta_t \approx 2+2d$, which could perhaps hold also for $d>3$. In 1d the SDRG results are confronted with numerical results obtained by free-fermionic techniques and a good agreement is found, which indicates that the SDRG results are possibly asymptotically correct in higher dimensions, too.

In 1d we have pointed out a relation between transverse-spin correlations in the RTIM and dimer correlations in the random AF XX-chain. This implies that the dimer correlations in the random singlet phase of the random AF XX-chain should have the same asymptotic form - including possible multiplicative logarithmic corrections - as that of the transverse-spin correlations in the RTIM in 1d. At this point we should mention that dimer correlations in the random AF Heisenberg (XXX) chain have been studied recently by Shu \textit{et al}\cite{shu}. According to SDRG arguments the IDFP-s of the random XX and the random XXX chains should be identical, thus the decay exponents of the dimer correlations of the two models should be the same. Indeed the numerical results about the random XXX-chain in Ref.\onlinecite{shu} obtained through numerical SDRG and by quantum MC simulations are consistent with our numerical findings on transverse-spin correlations in the RTIM in 1d.

Our results are expected to hold for other random quantum systems having discrete symmetry, such as for random quantum Potts\cite{senthil} and Ashkin-Teller models\cite{at}. Also connected energy-density-like correlations in stochastic models with strong enough disorder, such as the random contact process\cite{hiv,vd}, are expected to follow the same asymptotic decay as the transverse-spin correlations in the RTIM. These correspondences are expected to hold in higher dimensions, too.

\begin{acknowledgments}
This work was supported by the Hungarian Scientific Research Fund under Grants
No. K109577 and No. K115959. 
This publication was made possible through the support of a grant from the John Templeton Foundation. The opinions expressed in this publication are those of the authors and do not necessarily reflect the views of the John Templeton Foundation.
We thank to Yu-Cheng Lin for useful discussions.
\end{acknowledgments}

\section*{Appendix: Mapping of the XX-chain to two decoupled transverse-field Ising chains}

Here we recapitulate the well-known
mapping\cite{pst,IJR00,IJ07} of the Hamiltonian in Eq.(\ref{hamiltonian}) into two transverse-field Ising chains defined by
the Hamiltonians:
\beqn
H_{\sigma}&=&-{1 \over 4} \sum_{i=1}^{L/2} J_{2i} \sigma_i^x \sigma_{i+1}^x
-{1 \over 4}\sum_i^{L/2} J_{2i-1} \sigma_i^z\nonumber\\
H_{\tau}&=&-{1 \over 4}\sum_{i=1}^{L/2} J_{2i} \tau_i^x \tau_{i+1}^x
-{1 \over 4}\sum_i^{L/2} J_{2i-1} \tau_i^z\;.
\label{Hst}
\eeqn
Here the $\sigma^{x,z}_i$ and $\tau^{x,z}_i$ are two sets of Pauli
matrices at site $i$ and we have $H_{XY}=H_{\sigma}+H_{\tau}$.

One can easily find the transformational relations between the $XY$
and Ising variables:
\beqn
\sigma_i^x&=&\prod_{j=1}^{2i-1} \left( 2S_j^x \right),~~~\sigma_i^z=4 S^y_{2i-1} S^y_{2i}
\nonumber\\
\tau_i^x&=&\prod_{j=1}^{2i-1} \left( 2S_j^y \right),~~~\tau_i^z=4 S^x_{2i-1} S^x_{2i}\;,
\label{oprel}
\eeqn
whereas the inverse relations are the following:
\beqn
2 S_{2i-1}^x&=&\sigma_i^x \prod_{j=1}^{i-1} \tau_j^z ,~~~2 S_{2i}^x=\sigma_i^x\prod_{j=1}^{i} \tau_j^z
\nonumber\\
2 S_{2i-1}^y&=&\tau_i^x \prod_{j=1}^{i-1} \sigma_j^z ,~~~2 S_{2i}^y=\tau_i^x\prod_{j=1}^{i} \sigma_j^z\;.
\label{invrel}
\eeqn

\end{document}